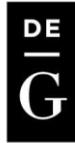

# BIODIVERSITY CONSERVATION AND STRATEGIES OF PUBLIC AWARENESS. CASE STUDY: THE NATURAL LANDSCAPES OF CENTRAL TUNISIA

## Islem Saadaoui [1], Christopher Robin Bryant [2], Hichem Rejeb [1], Alexandru-Ionuţ Petrişor [3]

**Abstract**. This research examines global issues concerning the development of mountain areas considered as territories difficult to manage. The case study area is part of the sub-region of High Alpine Steppes belonging to the Tunisian Ridge and reaching Tebessa Mountains in Algeria. The central question of this article is based on the analysis of the links between the representations produced by mountain landscapes and the construction of a border line that must meet the requirements of sustainable development. Eco-landscape determinants and the role of public authorities and population must be better defined so that the products of this space provide a better quality of life endowed with the alternatives of local and sustainable development. Our hypothesis is that the mountain areas of West Central Tunisia still have a real ecological potential little disturbed by a chimerical development, and can constitute assets for the territorial development of the area. The approach adopted by this work is a scoping audit based on the floristic richness and the monitoring of its spatiotemporal dynamics. The results of this research allowed us to draw rich conclusions; the phyto-ecology approach has shown a relative floristic richness that remains highly dependent on the climatic cycles and intervention of human action; this area must be considered as a priority of the public planning policies aimed at improving the quality of lives in these fragile zones in the context of sustainable development.

[1] PhD, Research unit "Horticulture, landscape an Environment", UR2013AGR06, ISA-IRESA,University of Sousse, Sousse, Tunisia
[2] PhD, Professor, Department of Rural Planning and Development, School of Environmental Design and Rural Development, University of Guelph, Guelph, Canada.
[3] PhD, PhD, Habil., Associate Professor and Director, Doctoral School of Urban Planning, "Ion Mincu" University of Architecture and Urban Planning, Bucharest, Romania.

## 1. Introduction

The evolution of human society has turned out to be extremely fast during the last centuries, producing extreme modifications of the natural environment, its well-defined footprint being represented by the anthropic environment with in depth modified characteristics (Petrişor *et al.*, 2010). The analysis of the existing relationship between human communities and the forest highlights the close correlation between the two basic subsystems of the geographical system (Guran-Nica *et al.*, 2008).

"*The forest biodiversity provides many ecosystem services, such as protection of plant, water, and soil resources*" (Rusu, 2012), and is also essential for maintaining the ecosystem's functionalities (Petrişor, 2015). According to Kemp (1992), forests are dynamics systems, their genetic diversity, particularly in forest formations relatively complex, is due not only to the number of species present in a given area, but also to the stages of succession.

Forests are also the cover of mountains which have always been special geographical, economic and social entities (Popescu and Petrişor, 2010). The relief, climate, history, and cultural heritage imply a specific policy of development, planning and protection. Much of the plant and animal species in Tunisia are located in mountain ecosystems (Saadaoui, 2017).

In addition, mountain areas are characterized by significant handicaps leading to more difficult living conditions and restricting the exercise of certain economic activities, in particular by a considerable limitation of the possibilities for land use and a significant increase in the costs of the work involved (Seibert *et al.*, 2001; Ianoş *et al.*, 2011).

Projects in landscape ecology cover a wide range of issues ranging from ecological risk analysis to the study of biodiversity and the development of socially acceptable management and spatial planning strategies as well as the observation of public attitudes towards landscape changes (Aghzar *et al.*, 2002); therefore, there is a need to protect the natural vegetation cover of arid areas and assess the environmental impact and economic cost (Matthew *et al.*, 2006; Kalpana *et al.*, 2007).

Biodiversity (the first link in trophic chains) is now recognized as a vital and common property. This article focuses on characterizing plant biodiversity in a mountain forest area and revealing its evolution in space and time. Our objective is to analyze the current dynamics of plant communities in order to be able to answer the following questions in the long run: do we observe dynamic stability or, on the contrary, are communities evolving? What are the underlying variables, and in particular what are the respective roles of natural (climatic) and local (anthropogenic) factors causing these changes?

Tunisia, part of the Mediterranean basin, is one of the richest regions of the world in terms of biodiversity of flora and fauna (Myers *et al*., 2000). Indeed, the Mediterranean countries hold almost 4.5% of the world's flora. Understanding the evolution of a space involves, first of all, analyzing the visual dimension (the landscape). As the first step in this article, we present and process the floristic data, and in particular the floristic lists of the years 2013, 2014 and 2015, in order to highlight the floristic composition and to characterize the floristic landscape of the study site.

Next, we have chosen to quantify the floristic biodiversity and to follow its spatiotemporal evolution through the calculation of the biodiversity indicators of the same site for three successive years. The final phase consists of statistical analyses using factorial analyses to reflect the effect of anthropogenic action on the distribution of flora in the mountainous regions of Central Western Tunisia.

## 2. Materials and methods
### 2.1. Study site

The study is conducted in three of the most representative municipalities in the region of Central West Tunisia (Fig. 1). These delegations are Feriana, Foussana and Hidra, which belong administratively to the Governorate of Kasserine, representing nearly 70% of the mountains of the Governorate and which share a 102 km borderline with Algeria (Commissariat Régional au Développement Agricole Kasserine, 1995).

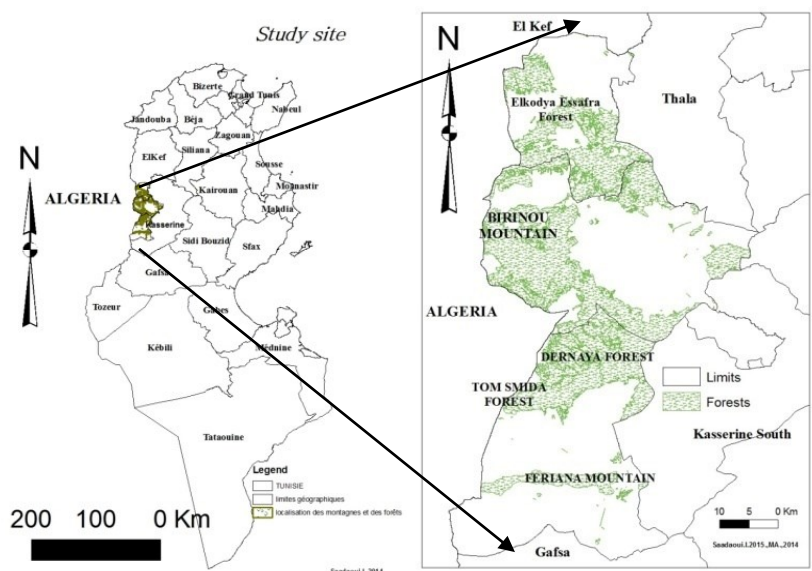

Fig.1. The geographic location of the study site (Saadaoui, 2014).

The study area presents part of the sub-region of high alpine steppe and covers an area of 235 thousand ha. The zone is located geographically between the two following coordinates:

- 8° 29'3'' N 35°40'13''E North.
- 8°26'30,53'' N 34°54'14,56''E South.

### 2.2. Methodology

The study attempts to inventory the natural vegetation and characterize it in ecological terms while highlighting the importance of environmental conditions and anthropogenic action. The method is based on a phyto-ecological analysis aimed at the quantification of the richness and the floristic diversity of the ecosystem of those forests using the Generalized Linear Modeling method, for determining the qualitative environmental realized niche (QERN) of plant species (Austin et al., 1990). Statistical analyses were carried out using the XLSTAT 2015 software to carry out the Factor Analysis of Matches (FAM) in order to represent the spatial dynamics in relation to the anthropogenic and natural conditions of the environment and ecological characteristics of flora. Statistical analyses were carried out to identify the spatial and temporal dynamics of the flora in relation to human activities: monitoring of frequencies and indicators of biodiversity (specific diversity, using the Simpson Index and Shannon Index) (Bryk et al., 1992; Gelman et al., 2007; Bolker et al., 2009; Scăunaşu et al., 2012; Lohbeck et al., 2017; Fontana et al. 2018; Kleinebecker et al., 2018; Mensah et al., 2018).

### 2.2.1. Sampling mode

The first stage is a prospecting phase because the study area was unknown to us, and it was carried out during the first visits during the year 2013. It was necessary to eliminate certain sites in order to choose those which, due to their originality and complexity, were considered appropriate to study.

The second stage is more practical; it is the operational phase in which we designed transects and plots, collected the plant samples, and surveyed the local populations both for natural factors and their activities. A comprehensive survey of species in the defined plots was carried out to determine the different biodiversity indicators for all the formations studied (Julve, 2000).

### 2.2.2. The surveys

We identified two parcels; the first is fenced and the second unfenced. Each has an area of 10,000 m². We placed ropes every 10 m in order to delimit the square plots of 100 m².

Species are counted and classified according to their biological and ecological characteristics. Obviously, we used the plots of 100 m² only when the wood groups were the most dominant. On the other hand, when the plant population is dominated by herbaceous plants, the minimum area used is 4 m² (2 m x 2 m) and a maximum of 25 m$^2$ (5 m x 5 m) depending on the environment studied (Muljosukojo, 1992).

### 2.2.3. Data processing

The floristic lists recorded on the site of the flora characteristic of the mountain zone are supplemented by quantitative and qualitative data which allowed us to characterize the flora and to describe the vegetation.

From these data, we assessed the spatiotemporal evolution of the annual and perennial flora of the site based on the computation of frequencies.

### 2.2.4. The supra-specific diversity

This index is important since it reveals alterations that could have been masked if we considered only specific richness. For example, a stand of about 100 species is considered to be less diverse if it is composed of species belonging to the same taxonomic group in contrast to the situation when they belong to very different groups. We estimated this supra-specific diversity at the level of gender and family (Lamotte, 1995).

### 2.2.5. Dynamics of the number of floristic species

The study assessed the dynamics of the number:

- of perennial species likely to contribute to the stability of the ecosystems of the arid and semi-arid zones.
- of annual species, sometimes characteristic of certain stages of succession.

In order to evaluate the significance of the variation in the state of the sample plots between protected and unprotected plots on the distribution of harvested species, we used a two-factor ANOVA test with experience replication, based upon the following hypothesis: "*the protection of forest plots has a powerful influence on the size of a species*".

$$FSC_i\% = \frac{n_i x 100}{N}$$

- Specific Frequency (SFi) of a taxon: number of points where this taxon was read
- The Centesimal Specific Frequency (CSFi) of a taxon: ratio in% between its FSi and the total number N of reading points along the line

## 3. Results

### 3.1. Current aspects of flora and vegetation

The present aspects of the flora are studied by treating the floristic lists resulted from a phyto-ecological survey on the two plots (fenced and not fenced) representative for the same type of territory. This analysis focuses on the nuances of floristic diversity and distribution of species in plant associations in response to different patterns of spatial organization and, therefore, to different degrees of anthropization of the distinct landscape (Ilahi, 2014).

We used the ecological data compiled from bibliographic sources and our field observations on the floristic lists.

### 3.2. Supra-specific diversity

Supra-specific diversity is revealed through the number of families and genera identified in the study site (Table 1).

Table 1. Number of families and genera.

|  | Family | Genus |
|---|---|---|
| **Observation site (Bouchebka forest)** | 17 | 41 |

### 3.2.1. Diversity of families

17 families are encountered in the study sites (Table 2); the main families are represented by 12 or more species which correspond mainly to annual species.

The two most represented families are *Asteraceae* and *Poaceae*.

The collected data included indicators of the presence of agro-pastoral activities. We also recorded the level of crops, weeds and, grasslands. We distinguished a broader distribution of grasses (*e.g.*, quackgrass, common barley, and plantains) in the unfenced plot compared to only 27% of the species in the fenced plot. *Plantaginaceae* are distributed mainly in the unfenced plot with a rate of 81%. These species are indicators of high anthropogenic activity, and represent weeds and wastelands.

Table 2. Number of species by census families.

| Families | Number of species | Families | Number of species | Families | Number of species |
|---|---|---|---|---|---|
| 1. Asteraceae | 12 | 7. Plantaginaceae | 6 | 13. Myrtaceae | 1 |
| 2. Pinaceae | 1 | 8. Fabaceae | 4 | 14. Mymosaceae | 1 |
| 3. Cupressaceae | 3 | 9. Lamiaceae | 4 | 15. Caryophyllaceae | 1 |
| 4. Oleaceae | 1 | 10. Thymeleaceae | 2 | 16. Apiaceae | 2 |
| 5. Cyperaceae | 1 | 11. Brassicaceae | 3 | 17. Cistacea | 1 |
| 6. Poaceae | 15 | 12. Linaceae | 1 |  |  |

We also note an abundance of *Asteraceae* with rates of 43% in the fenced plot and 57% in the unfenced plot (Fig. 2).

*Pinaceae* and *Cupressaceae* have higher rates in the fenced plot; this distribution was caused by the deforestation of tree species in the unfenced plot.

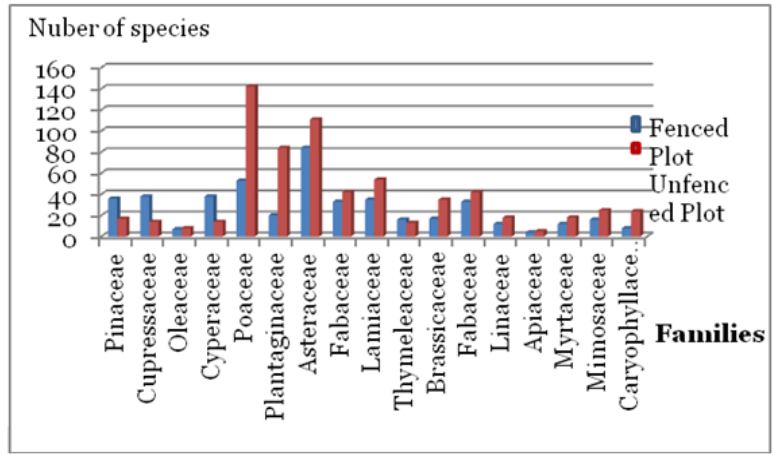

Fig. 2. Distribution of the different families by number of individuals.

### 3.2.2. Genus diversity

The flora of the study area is composed of 41 genera. The majority of the genera listed are represented by a single species, while 21.9% of genera are represented by more than one species.

## 3.3. Description of vegetation: the vegetation spectrum of Tunisian central west mountains

This phase of the floristic study was devoted in the first stage to the characterization of the global vegetation within the study area. Then, an analysis of the distribution of these species according to their autecological characteristics influencing their distribution on the site was carried out.

### 3.3.1. Distribution of flora according to intrinsic characteristics

Adventitious species had the highest rate, 35%, while species that reflect anthropogenic activity had a rate of 44%: 24% for prairie species, 10% for cultivated species and 10% for wasteland species.

This distribution indicated the presence of strong anthropogenic action in the study site (agrpastoral activities).

In order to refine this analysis, we compared the distribution of the number of species in the two different neighboring plots (Fig. 3), which guided us to the following results:

- Weeds, cultivated and grassland species are predominantly distributed in the unfenced plot,
- We note a much smaller distribution of the cultivated species in the fenced plot.

The analysis of the autecology data showed that the study site is an environment characterized by its drought. More than 70% of the species encountered are species that adapt to poor or very poor water soils. The species found in the study site are very demanding in relation to light. This indicates a dense woody environment, allowing the different species to benefit from the light, and consequently a strong distribution of the heliophilic species.

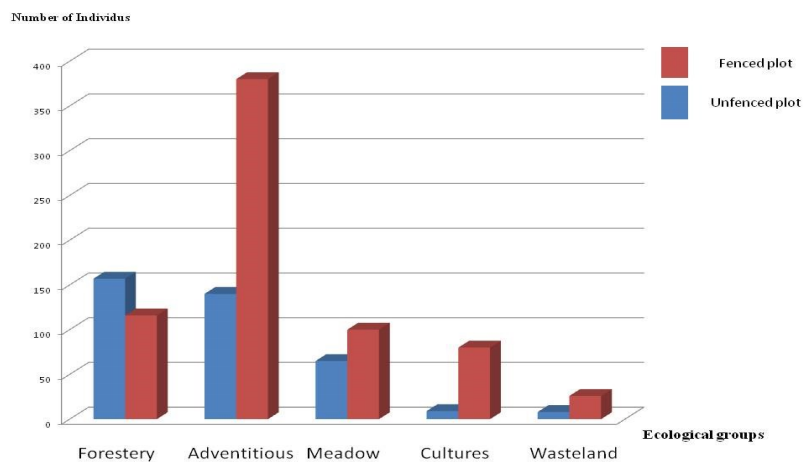

Fig. 3. Distribution of ecological groups of flora per plot.

Woody plants represent respectively 36% of the species encountered in the study site. This demonstrates that the environment is a very dense woody environment. Hemicryptophytes, therophytes, and geophytes account for 33%, 24% and 7%, respectively, of the vegetation of the study area. This leaded to the conclusion that the flora is mainly represented by annual summer species and perennial herbaceous plants.

The study of the modes of dissemination of the flora (Fig. 4) led to the conclusion that those zones are characterized by their windy aspect and arid climate.

The species which are disseminated by anemochory are the most encountered, with a rate of 35%. This mode of dissemination suggests that this region is distinguished by its windy appearance. We encounter this mode of dissemination not only in annual and biennial species but also for Aleppo Pine species. This forest tree, which dominates the tree flora of the region, is also disseminated by means of anthropochory (regenerated). 5% of the tree species encountered, such as *Eucalyptus* and *Acacia*, are regenerated. Species disseminated by epizoochory also have a high rate (21%). Species disseminated by barochory have a rate of 32% among the species encountered.

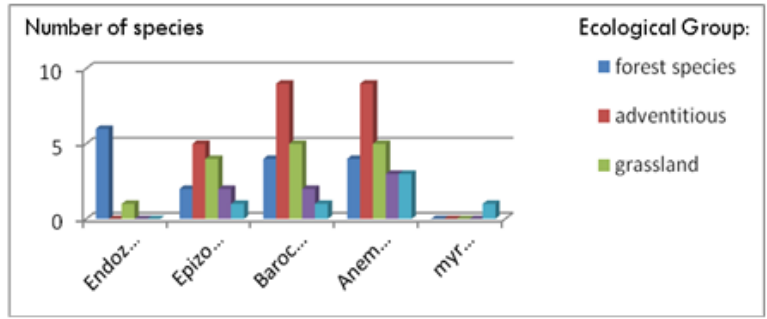

Fig. 4. Modes of dissemination.

### 3.3.2. Spatiotemporal analysis of the flora of the study site

*3.3.2.1. Evolution of the numbers of floristic species.* Based on the plant data, we noticed a greater number of perennial species in the fenced plot, with the most important species for the year 2013 being the woody ones.

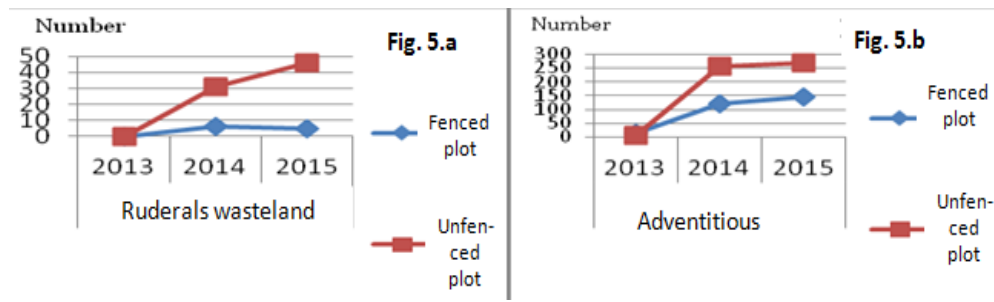

Fig. 5. Spatiotemporal evolution of weed and grassland numbers.

We noticed an evolution in the number of fallow species that were totally absent during the year 2013 which then registered 5 species of wasteland, compared to the year 2015 with 59 individuals (Fig. 5a). Weeds also showed a significant increase in the number of species and of individuals by species (Fig. 5b). These species were mostly present in the unfenced plot where the anthropogenic action is stronger.

Provided the increase of precipitations (rainfall and snowfall) recorded for the years 2014 and 2015, we noticed a considerable increase in the number of species but also with the appearance of new species during the sampling.

As expected, the distribution differed according to the type of the plot. Other than the effect of natural factors, the distribution of cultivated and grassland species depends on the presence of anthropogenic activity, which indicates a more intense distribution in the unfenced plot. The study area is very close to an urban agglomeration, where agriculture (family farming) and pasture (grazing) activities have been noticed.

Associated with the natural factors of the study area, anthropogenic activities are responsible for the distribution of cultivated, grassland and wasteland species in the study sites.

For forest species, the mountainous zone of the Center-West region of Tunisia is characterized by the abundance of the Aleppo Pine forest species. We noticed a decrease in the number of forest species in the 2014 surveys. This decrease is more visible in the unfenced plot. It is mainly due to the effect of the deforestation carried out by the inhabitants of this mountain area and the total absence of enforcement by agents of the Directorate General of Forests.

A slight increase in the number of trees is noticed for the year 2015. During the field visits, we noticed the presence of young development of Aleppo Pine regenerated by seeding, *Eucalyptus* and also of *Acacia*.

Deforestation remains the most serious damage to the forest ecosystem. More than three hundred feet of Aleppo pines are cut off each year, and their wood is sold out to public baths and other operators.

*3.3.2.2. The frequency of annual and perennial species and their evolution over time.* The analysis of the spatiotemporal variation of the flora yielded the following results:

The number of perennial species decreased from one year to the next. We noticed a decrease of the specific centesimal frequency in the two parcels:

- In the fenced plot, from 76.8% for the year 2013 to 39.1% for the year 2015
- In the unfenced plot, from 47.5% to 17.2%.

Perennial species are deforested by the population of the area, especially within the unfenced plot, but this action is also carried out in the protected plot, suggesting the existence of a major threat to the forest ecosystem of the area. The analysis of the evolution of the number of annual species showed a significant increase in the specific frequency throughout the territory from the year 2013 to the year 2015.

In order to study the dynamics of evolution in time and space of perennial and annual species, we used the "ANOVA test" for two factors with repetition of experience.

The results are presented in Tables 3 and 4.

### The case of perennial species

There was a significant variation in the distribution of species between the different sampling plots since the probability P is highly less than 0.05.

Indeed, when analyzing the interaction within the plots, we noticed that there is no significant difference in the distribution of perennial species for the years 2014 and 2015 within the same plot (Table 3).

Table 3. Analysis of variance: two factors with repetition of experience – the case of perennials.

| DETAILED REPORT | Fenced plot | Unfenced plot | Total |
|---|---|---|---|
| *Perennial species in 2014* | | | |
| Number of samples | 14 | 14 | 28 |
| Variance | 68,7472527 | 24,5934066 | 57,6455026 |
| *Perennial species in 2015* | | | |
| Number of samples | 14 | 14 | 28 |
| Variance | 64,7472527 | 46,9945055 | 58,0992063 |
| *ANOVA Two factors with repetition of the experience* | | | |
| *Variation sources* | *F* | *Probability* | *Critical value for F* |
| Sample | 2,06502344 | 0,1567027 | 4,02663122 |
| Columns | 8,36771601 | 0,00556619 | 4,02663122 |
| Interaction | 0,5854789 | 0,4476317 | 4,02663122 |
| Inside the group | 2108,5 | 86 | 24,5174419 |

Table 4. Analysis of variance: two factors with repetition of experience – the case of annual species.

| DETAILED REPORT | Fenced Plot | Unfenced Plot | Total |
|---|---|---|---|
| *Annual species in 2014* | | | |
| Samples number | 43 | 43 | 86 |
| Variance | 20,6234773 | 58,3089701 | 47,4155951 |
| *Annual species in 2015* | | | |
| Samples number | 43 | 43 | 86 |
| Variance | 23,0863787 | 50,4518272 | 45,2318741 |
| Analysis of Variance (ANOVA) | | | |
| *Variation sources* | *F* | *Probability* | *Critical value for F* |
| Sample | 2,41914247 | 0,00027412 | 1,52588238 |
| Columns | 59,9973915 | 1,723111 | 3,95188225 |
| Interaction | 1,78761702 | 0,01181139 | 1,52588238 |
| Inside the group | 2108,5 | 86 | 24,5174419 |

**The case of annual species**

For the annual species, we did not notice a significant difference in the distribution of the species numbers over time (between 2014 and 2015) for the two sampling plots (closed plot and unfenced plot). This shows that the natural and anthropogenic conditions between the two years of monitoring did not have a strong influence on the distribution of flora in the study area.

Furthermore, we noticed that P is far away lesser than 0.05 when monitoring the evolution of the number of species within the same plot. We can conclude that the plot protection factor can influence the distribution of annual species. This is reflected by a higher number of species in the fenced plot, since the population practices husbandry mainly in the unprotected areas (Table 4).

### 3.4. Biodiversity quantification
### *3.4.1. Hierarchy of factors influencing vegetation distribution and floristic composition*

**In summary for the overall floristic list:**

By interpreting the factor plane F1-F2 in Fig. 6, it can be seen that at the level of the positive coordinates of the F1 axis we found the taxa of species with a steppe aspect, such as *Hyparrhenia hirta*, *Stipa tencissima* and *Stipa capillata* etc.

These species showed a positive trend towards the formation of steppe and lean grasslands, in relationship to the steppe and the pastoral character of the site.

The site was distinguished by a distribution of a significant number of forest species (*e.g.*, *Pinus halepensis*, *Juniperus oxycedrus*, *Phillyrea latifolia*, *Eucalyptus camaldulensis*, *Acaciamangium Willd*, *Cistus creticus*, and *Scirpus sylvaticus*). We noticed their distribution on the positive side of the factorial axis F1. The F1 factor design showed a positive interaction between steppe species and forest species.

The same plan showed a negative interaction between cultivated and ruderal species with the forest ones; this interaction indicated the presence of anthropogenic action in a forest environment.

These results suggested the forestry and pastoral vocation of the study site.

**The negative coordinates of F1 axis**

There was a cultivated appearance (*Avena sterelis*, *Hordeum vulgaris*, *Artemisia vulgaris*, *Othonna cheirifolia*) with a remarkable distribution of adventitious species, which indicated the presence of anthropogenic action.

**For the F2 axis**
**Positive coordinates**

Only the forest species (*e.g.*, *Pinus halepensis* and *Scirpus sylvaticus*) hade a high positive index, but these species are negatively correlated with the steppe and prairie species (*e.g.*, *Stipa tenacissima*, *Thymelaea hirsuta*, and *Retama retam*). The distribution of these species is influenced by anthropogenic actions.

The site is a forest environment, a nature influenced by human practices (forest environment correlated with the steppic and prairie vocation).

**Negative coordinates**

Adventitious, cultivated and uncultivated species had negative indices, suggesting the existence of weak anthropogenic actions in the study area. These species are positively correlated with grassland species.

We concluded that anthropogenic activities favored the distribution of prairie species. It is mainly the pasture and the contribution of the herds in the site that favors the dissemination of certain species (dissemination by endozoochory and epizoochory).

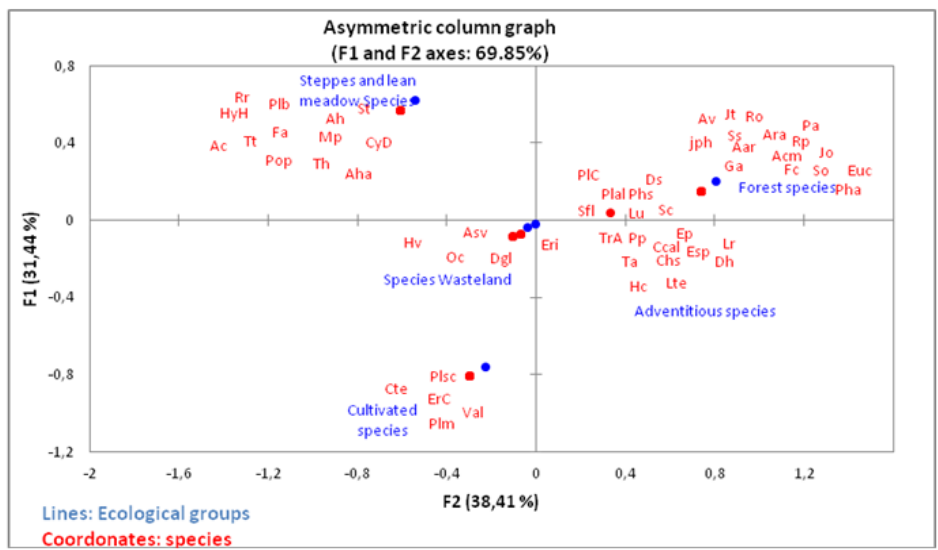

Fig. 6. Interaction of species by ecological groups in the study site.

**Conclusions**

Through this article and based on the phyto-ecological approach, we have reached our objectives: the description and understanding of vegetation, and the two-dimensional spatial and temporal organization, qualitatively and quantitatively, of the plant species that constitute it. The Centre West of Tunisia is an area of the arid steppe with xerophilic formations adapted to low precipitation (150-400 mm/year), high thermal amplitudes, intense evaporation and a high frequency of dry winds. In this article, we have studied a forest environment with a steppe tendency with a clear presence of anthropogenic actions, given the geographical location of the study site, its proximity to an urban agglomeration, presence of farms close to the forest environment, all justifying the presence of fallow and cultivated species in the studied plots.

Despite the fragility of these landscapes generally linked to the physical conditions of the natural environment and pressures exerted by the anthropic factor, the development of this area is always possible through a landscaping mediation that connects the forces of decision-makers, managers and involves the occupants of this region who could switch (we hope) to a potential and formal sustainable development of their territory. On the same note, it is necessary to draft overall management plans that safeguard the sustainability of the human and non-human landscape elements.

It would be wise to continue this cross-disciplinary research to detect the real landscape potential of the border areas and eventually propose concrete actions of eco-development.